\documentclass[preprint,authoryear]{elsarticle}
\usepackage{graphicx}
\usepackage{natbib}

\title{A large HPGe detector for the non-destructive radioassay of an ultra-low-background counting facility}
\author[max]{M.~Fechner\corref{corr1}}
\ead{maximilien.fechner@cea.fr}
\author[bob]{C.~Henson}
\author[jon]{J.~Gaffiot}
\author[max]{T.~Lasserre}
\author[jon]{A.~Letourneau}
\author[jon]{D.~Lhuillier}
\author[max]{G.~Mention}
\author[jon]{Th.~A.~Mueller}
\author[max]{R.~Qu\'{e}val}
\author[bob]{R.~Svoboda}
\address[max]{Commissariat \`{a} l'\'{E}nergie Atomique et aux \'{E}nergies Alternatives, Centre de Saclay, IRFU/SPP, 91191 Gif-sur-Yvette, France}
\address[jon]{Commissariat \`{a} l'\'{E}nergie Atomique et aux \'{E}nergies Alternatives, Centre de Saclay, IRFU/SPhN, 91191 Gif-sur-Yvette, France}
\address[bob]{Department of Physics, University of California, Davis, Davis, CA 95616, USA}
\cortext[corr1]{Corresponding author}

\begin{document}

\begin{abstract}
 We present the use of a low background counting facility, equipped with a p-type 80\% 
relative efficiency HPGe detector, protected by active and passive shielding, and large enough to count
a 10'' photo-multiplier tube (PMT). A GEANT4 Monte-Carlo of this detector was developed and tuned to
3\% accuracy. We report the U, Th, and K content in three different types of PMTs
used in current neutrino experiments, with accuracies of $\sim 10$ ppb for U and Th
and of $\sim 15$ ppm for K.
\end{abstract}

\begin{keyword}
 HPGe detector \sep gamma spectrometry \sep low background \sep Monte-Carlo simulation \sep photomultiplier tube.
\end{keyword}

\maketitle

\section{Introduction}

 The aim of this paper is to present the use of a High Purity
 Germanium gamma-ray detector (HPGe) in an underground counting room at
 Saclay near Paris, France, along with a Monte-Carlo simulation and analysis 
 software suite. This low-background facility is a joint project between the University of California
 at Davis and the Commissariat à l'Energie Atomique et aux Energies Alternatives 
 (French Atomic Energy and Alternative Energies Commission or CEA).
 The counting room is
 being used by the Double-Chooz~\citep{DChooz} and Nucifer~\citep{Nucifer}
 neutrino detection projects, whose low radioactivity background
 requirements warrant precise radioassays of all detector building materials.
 Both detectors use large volumes of liquid scintillator and detect light pulses
 with large area photo-multiplier tubes (8 or 10 inch tubes).
 The large
 inner volume (27 cm diameter by 38 cm height) of the lead shield surrounding the HPGe detector was used to
 radioassay large samples which could not fit in other counting rooms
 from collaborating institutes, especially photo-multiplier tubes (PMTs).
 Furthermore, because of its closeness with the Chooz site (Ardennes, France),
 this counting room was used extensively during the detector's construction. 

 We will first present a detailed description of the counting room's
 set up, and then show the details of the development and tuning of a
 GEANT4-based Monte-Carlo simulation~\citep{Agostinelli:2002hh} of the HPGe
 detector. The simulation was used to predict the
 gamma-ray detection efficiency at all energies for a variety of
 geometries. 
 Finally we will detail the simulation
 and radio-assay results of three photo-multiplier tubes (PMT),
 used in Double-Chooz, Nucifer and Antares~\citep{Antares}. The simulation allows a complete description of
 the geometry of complex objects, improving the accuracy of the
 radioassay. In turn these measurements were used to predict the
 background from radioactivity in the neutrino detectors, improving
 their sensitivities.

\section{Description of the detection system}

\subsection{HPGe crystal}
\begin{figure}[!htb]
\begin{center}
\includegraphics[scale=0.5]{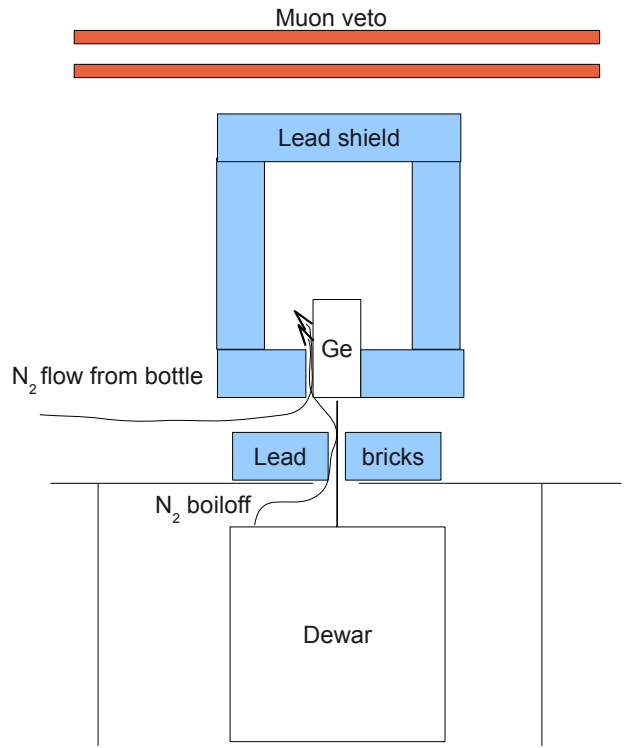}
\caption{Low-background setup including shielding and detector.}
\label{fig:geom}
\end{center}
\end{figure}

   The HPGe detector is a p-type coaxial detector of 80\% relative efficiency,
   manufactured by ORTEC.
   It is mounted on a vertical cryostat, 30 cm directly on top of a nitrogen dewar. 
   The crystal and its preamplifying electronics are enclosed in
   a high-purity copper endcap, and the crystal is protected by a commercial
   lead shield from ORTEC described below. Note that the position of the crystal
   within the shield is not fixed: it can be moved up and down to increase
   the available space in the counting chamber. In this work the crystal was placed
   as low as possible, with the bottom edge of the endcap flush with the
   bottom of the shield (see figure~\ref{fig:geom}). This configuration is non standard
   and increases the amount of background since the shielding is less efficient,
   but it allows us to fit a 10'' PMT inside the shield.
   
   The HPGe detector is connected to an ORTEC DSPEC jr 2.0 digital spectrometer,
   and ORTEC's MAESTRO 32 program is used for data acquisition. Spectra are then 
   translated to ASCII and processed with custom-made programs based on ROOT\citep{ROOT}. 
   Automatic pulse shaping and pole-zero correction settings were used, and the energy
   scale was calibrated using radioactive sources.

\subsection{Underground laboratory at Saclay}
The whole counting facility is placed underground in an alcove of an old
experimental hall at the Saclay center of the Commissariat \`{a} l'\'{E}nergie
Atomique et aux \'{E}nergies Alternatives (French Atomic Energy and
Alternative Energies Commission or CEA) near Paris, France. This
underground hall previously housed an electron accelerator
(Acc\'{e}l\'{e}rateur Lin\'{e}aire de Saclay or ALS). After its
decommissioning in the 1990s, the hall was cleared and is no longer considered a
radioactive area. However some of the concrete in the walls and ground
has a slight residual activity. The impact of this ambient noise has
been studied, and reduced as much as possible with lead shielding.

The main advantages of this experimental hall are in its availability at no cost,
great proximity and ease of access for our team. In addition, it has a fairly
large concrete and soil overburden, reaching at least 15 meters of water equivalent (mwe)
in all directions.

\subsection{Background rejection methods}

 Unshielded measurements with a small, portable germanium counter inside the hall revealed 
 the main $\gamma$-ray lines to be $^{152}$Eu and $^{60}$Co. 
These two elements have been shown~\citep{russians} 
to be produced by activation of concrete, which did occur to some extent when
an accelerator was in use in this hall. The walls also show strong $^{40}$K and $^{208}$Tl
activity.
In order to protect the spectrometer from this ambient noise, a lead shield from ORTEC
was used.
This lead shield is cylindrical, and 10 cm thick, with an inner 3-mm copper lining.
It protects the crystal from ambient gamma radiation on all sides, except
the bottom (see figure~\ref{fig:geom}), which is therefore sensitive to gamma rays 
from the ground and cryostat.
In order to reduce this background, two layers of lead bricks were placed on
top of a table located between the cryostat and the crystal.
The table and lead shield have circular holes, allowing the cryostat to poke through (see figure),
and the bricks were arranged so that only a small gap around the cryostat is left
unshielded.

 Since our shield is not airtight, radon gas is a troublesome source of background, 
as $^{214}$Pb and $^{214}$Bi cause many strong gamma lines in our spectra. 
In order to reduce this background,
the lead shield was lined with 5-mm thick rubber insulation along
its main openings. The bottom of the shield was covered with a sealed plastic sheet,
and the boil-off nitrogen gas from the dewar was directly piped into the shield.
A small permanent flow of nitrogen gas at 190 mL/min is constantly maintained
using bottles. 
With these measures we observe a reduction of a factor of roughly 7 in radon activity,
however radon gas remains a significant background with typically
$10^{-2}$ counts/s in the 609 keV $^{214}$Bi line.
The detector, cryostat and electronics are also kept inside an ISO 7 clean tent to reduce
the amount of dust that enters the experimental area.

\begin{figure}[htb]
 \begin{center}
\includegraphics[scale=0.5]{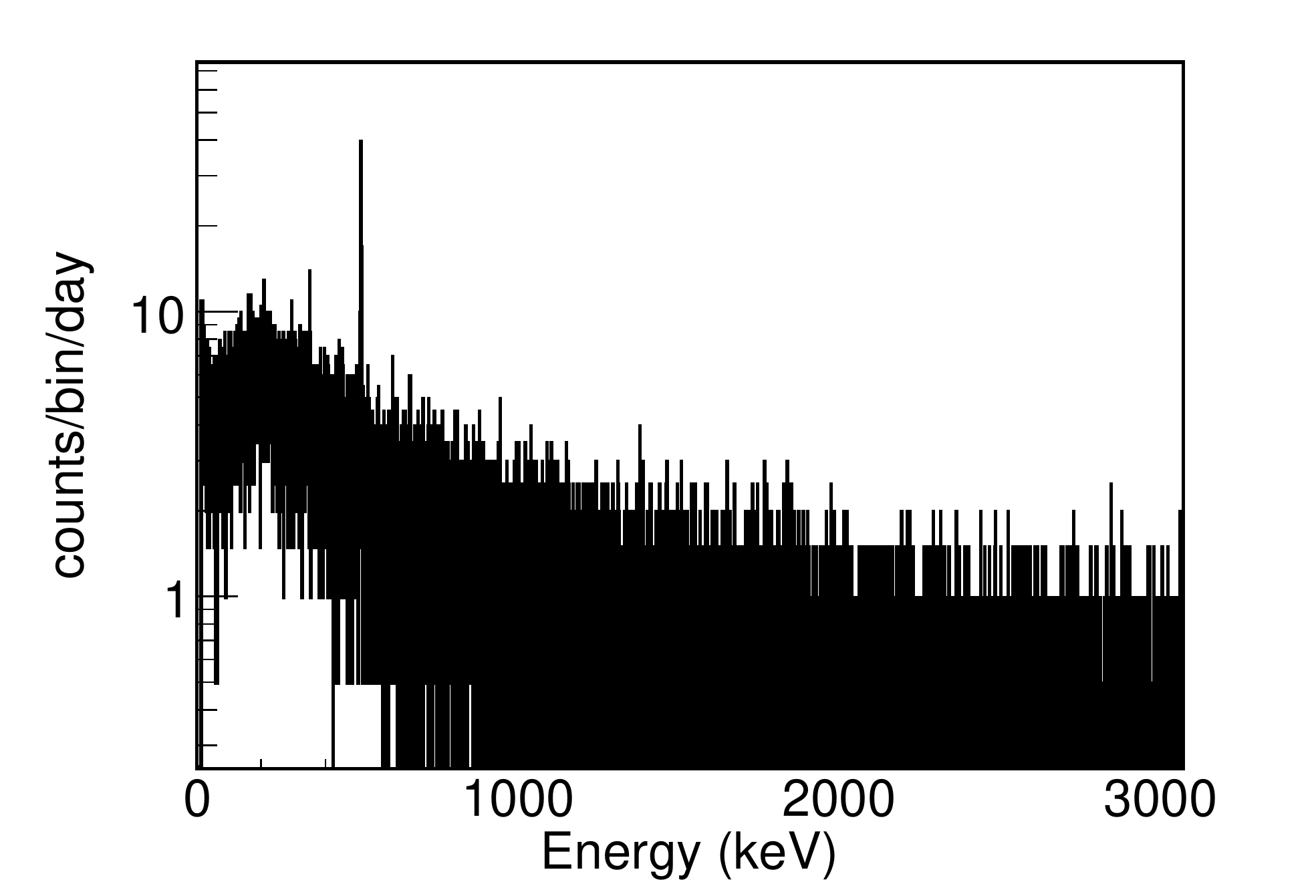}
\caption{Spectrum obtained in coincidence with the muon veto. The bin size is 0.19 keV and the run time 
for this acquisition was 2.02 days. 
The characteristic 511 keV peak of positron annihilation is clearly visible and enhanced, showing selection
of muon induced events.}
\label{fig:bkgcoinc}
\end{center}
\end{figure}

\begin{figure}[htb]
 \begin{center}
\includegraphics[scale=0.5]{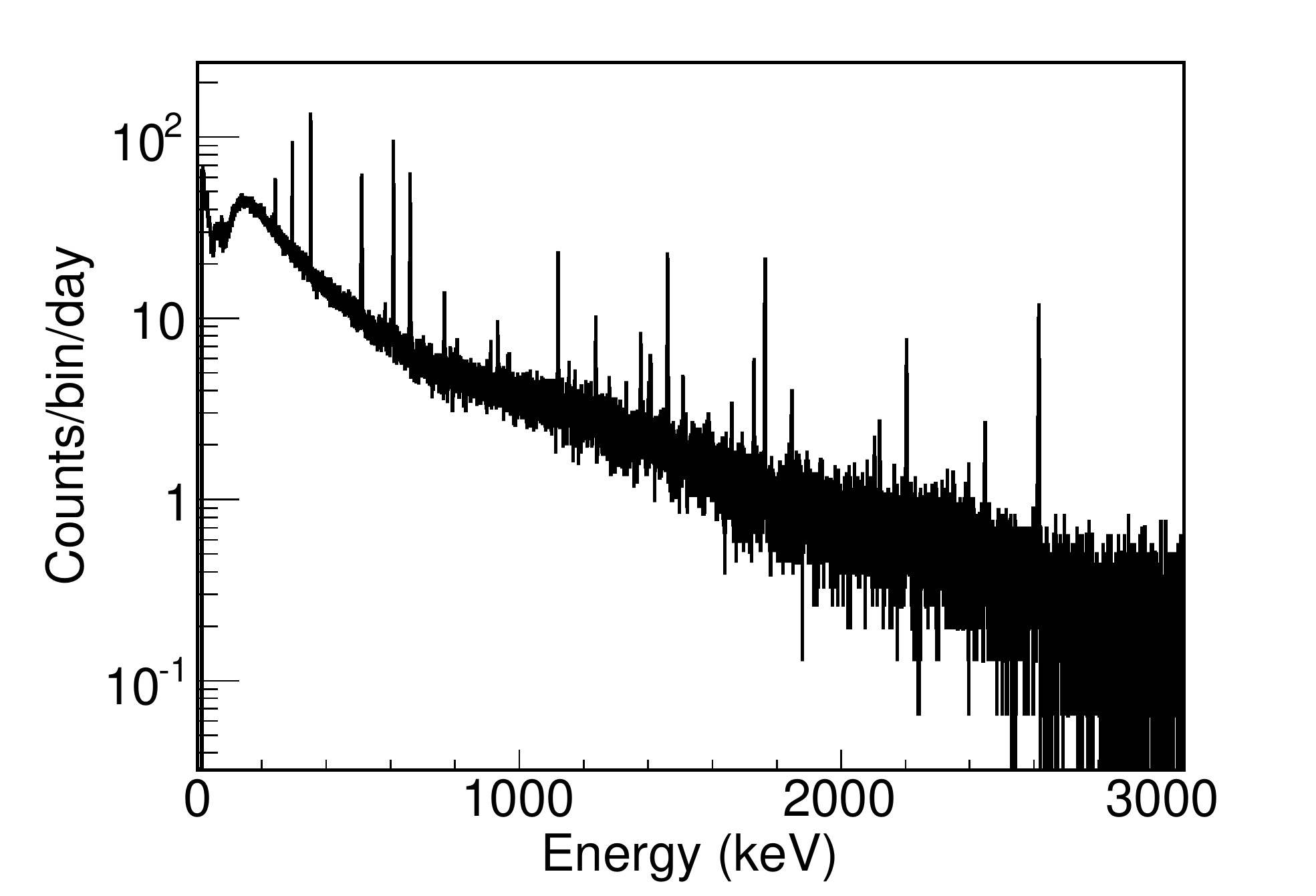}
\caption{Background spectrum obtained in July 2010. The bin size is 0.19 keV and the run time 
was 17.61 days.}
\label{fig:bkg}
\end{center}
\end{figure}

 Finally, the cosmic ray contribution is reduced by an active
veto made of two plastic scintillator panels, $39\times39\times1$ inch thick. These two
panels are both laid on top of the lead shield, centered
around it, one above the other. They are each viewed by a pair of 2-inch PMTs. The output of each PMT
is sent to a constant fraction discriminator, and the OR of the discriminated
signals is used to generate a long TTL gate with a Lecroy 222 gate generator.
Pulses from the HPGe detector occurring a few $\mu$s after the veto was triggered
are vetoed, taking only those in anti-coincidence 
with the TTL gate. 
The veto's average trigger rate is roughly 600 Hz, leading to 0.6\% dead-time.
The observed reduction between veto on and veto off is a factor of about 1.8 on the 511 keV peak
(see figures~\ref{fig:bkgcoinc} and \ref{fig:bkg}) but the overall effect on the spectrum is small.
Improving the light collection system could lead to a higher muon detection efficiency 
and therefore more efficient vetoing, consistent with what has been reported by many authors.

With all these rejection techniques, our typical background spectrum obtained with an empty 
counting chamber is shown in figure~\ref{fig:bkg}.

The overall shape is characteristic of muon-induced background~\citep{semkov},
with several prominent X and $\gamma$-ray lines.
 The main lines that can be observed are those of $^{222}$Rn daughters 
($^{214}$Pb and $^{214}$Bi) as well the annihilation line at 511 keV,
caused by cosmic rays. Below 100 keV, Pb-X-rays are strongly reduced by
the copper lining of the shielding, at the cost of increasing the background
at higher energies~\citep{canberra}. $^{40}$K and $^{208}$Tl are also
clearly visible.
A small contamination by $^{137}$Cs can
also be seen in this figure. This comes from the use of contaminated lead bricks
in an experiment in the hall, which were promptly removed, but left small 
trace contamination.
Our background rate varies from 0.2 to 0.3 counts/second/100 cm$^{3}$ of detector 
between 50 and 2700 keV, for a counting chamber with a total volume of about 22 L.

\section{Analysis method}

A dedicated spectral analysis program was developed for this
analysis. It is written in Python and uses ROOT~\citep{ROOT} to perform fits and 
histogram-based calculations.
First, a peak finding algorithm identifies good peak candidates based on statistical
and goodness of fit criteria. This algorithm is based on the one from GRABGAM
described in \citet{Winn2000430}.

  For each identified peak or user-input region of interest (ROI), gross counts
and net counts are calculated, as well as their respective errors.
To obtain net counts, continuum subtraction is performed 
using the formulas from \citet{Morel19981251}, which take into account a smoothly
varying background with different heights before and after the peak.
Using a database of X- and gamma-ray peaks, each peak is identified.

  When performing an actual radioassay, the sample's spectrum and a ``blank spectrum''
(taken immediately before or after the measurement) are compared, 
and statistical calculations are carried out to extract either
an upper limit or a measurement on each relevant gamma ray line. 
The results are presented as in GRABGAM~\citep{Winn2000430},
but the usual \citet{currie} minimum detectable activity (MDA) 
and decision thresholds are also computed.
The program is fully automatic and produces a report in PDF format.

An important input for accurate results is the detection efficiency as a function
of gamma ray energy, which we obtain with the Monte-Carlo simulation described in
the next section.

\section{Tuning the Monte-Carlo simulation}

  The subject of HPGe Monte-Carlo tuning is an active topic in the literature. 
The basic technique is always the same: using a Monte-Carlo simulation
software tool, the geometry and materials of the germanium crystal and all
surrounding layers are coded. At first, the manufacturer's dimensions are used,
but they are usually not known precisely enough, leading to large data/simulation
discrepancies. Using measurements with various point sources, the simulation is tuned by varying
several geometry parameters until satisfactory agreement with data is reached.
For our material screening purposes, Monte-Carlo accuracy of under 5\% 
is considered sufficient.

 Among the extensive literature on this topic, we have tried to follow
the method of \citet{Budjas2009706}, outlining the tuning of a smaller p-type
detector similar to the one we use.

\subsection{Implementation of the geometry in the Monte-Carlo simulation}
  We used the GEANT4~\citep{Agostinelli:2002hh} simulation package developed at CERN
for our Monte-Carlo (MC) simulation. Users are required to precisely describe the geometry
and material of each detector component in the program's code. The package
can then calculate the detector's response to the passage of particles,
propagating each particle step by step and taking into account all 
physics effects, including secondary particle production.

  The active detector is simply a volume of pure germanium.
  The detector response is simulated using the measured FWHM as a function of
energy, and is assumed to be Gaussian.
The energy deposited in the active volume by ionizing particles is added up, and then smeared
using the measured resolution. All MC peaks therefore have Gaussian shapes,
which is only approximately correct.

  A critical part of the simulation is the geometry of the detector,
  its mount cup and end cap, as well as any other material that stands
  between the crystal and the source.  Several authors,
  e.g. \citet{Hardy200265}, explain how they described the bulletized
  edge of their germanium crystal by approximating the detector with
  stacked cylinders of varying radii. 
  In this work we chose to describe the germanium crystal as a Boolean
  union of two cylinders and one torus. The bulletized edge's radius
  of curvature (small radius of the torus) was set to 0.8 cm, as 
  reported on the manufacturer's data sheet.

\subsection{Dead layer tuning}

  The ``dead layer'' is an inactive germanium layer on the surface of p-type
crystals, usually about 700 $\mu$m thick, caused by the drifting of Li ions
to dope the semi-conductor. Its thickness has a strong effect on the detector's
efficiency below a few hundred keV and is the main tuning parameter in our
simulation. 
  
  In order to describe the dead layer, a ``daughter volume'' of the germanium crystal
with the same shape, but slightly smaller, adjustable dimensions was implemented 
and declared to be the active volume. The small outer region of the germanium
volume which does not belong to the active daughter volume is the dead layer of
the simulated detector.
The main advantage of this method is that its dimensions can then be easily 
changed without complicated geometry calculations.

 Using a 3.33 MBq $^{241}$Am-Be source, we tuned the thickness of this region,
 following the method outlined in \citet{Budjas2009706}.
 The source itself was not precisely calibrated, but the method uses the ratio
 between several lines, eliminating the need for precise source calibration. 
 The source was placed on a custom made holder, 26.0 cm above
 the center of the detector's endcap, and left to count for a few hours 
 to accumulate enough statistics for a negligible statistical error.
 The dead time recorded by the spectrometer ($<10\%$) was
 deemed small enough to trust the measurements. No significant
 peak-sum effect was found.

 We measured the full-energy peak efficiency of the 59 keV line.
 Our detector having good enough energy resolution, we could
 distinguish a multiplet with four lines, at 96 keV, 99 keV, 101 keV, 
 and 103 keV, the two lines at 99 keV and 103 keV being the strongest. 
 In order to avoid systematic effects coming from line
 separation, we used a region of interest (ROI) spanning all four
 lines, from 95 keV to 105.5 keV, and performed continuum subtraction
 on the whole multiplet.
 We computed the efficiency ratio between the 59 keV line and the multiplet,
 $R(\mathrm{multiplet}/59)_{\mathrm{data}}$, 
 to cancel the systematic error from absolute source calibration as 
 outlined in \citet{Budjas2009706}. 

 Since we were interested in gamma ray lines below 150 keV, any material
 located along the path of the photons had to be taken into account. We carefully implemented
 the 5 mm thick plastic source housing, with two 1 mm thick Be sheets surrounding
 the 11 mm diameter Am deposit. The source was laid flat on top of an adjustable, 2 mm thick 
  plastic holder, which was also measured with calipers and implemented in the simulation.
 Having observed wrong results in the GEANT4 simulation of the gamma lines associated
 with $^{241}$Am decays\footnote{This has also been reported in \citet{Budjas2009706},
  and has apparently been fixed in later versions of GEANT4. Other elements are unaffected
  and are in agreement with ENSDF.}, we implemented our own Am generator, using the main X and gamma lines
 from ENSDF databases, and neglecting all cascade and angular correlation effects 
 (such effects are negligible considering the very small branching ratios of all lines above 59 keV).
 
\begin{figure}[!htb]
\begin{center}
\includegraphics[scale=0.4]{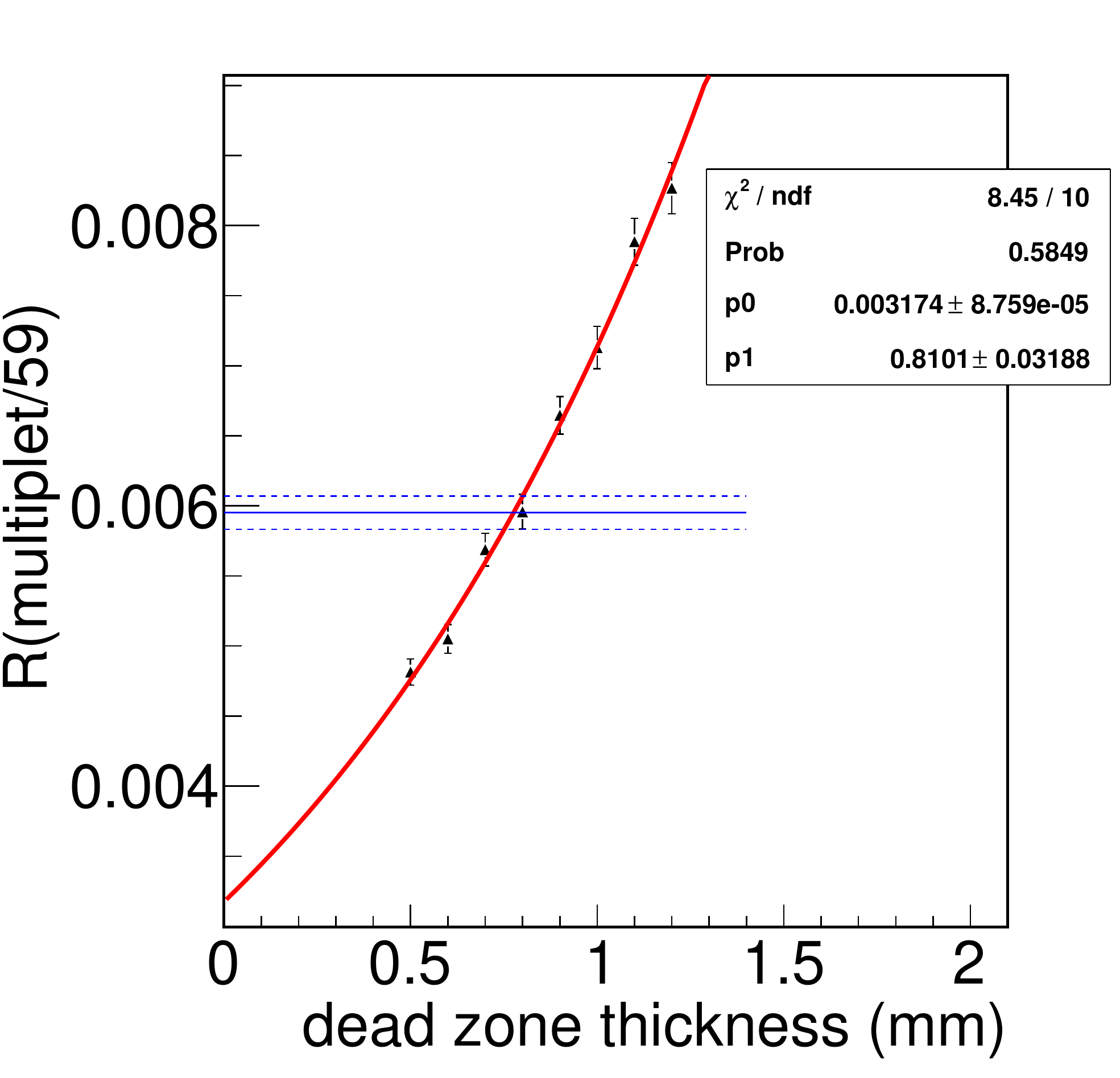}
\caption{Variation of $R(\mathrm{multiplet}/59)_{\mathrm{MC}}$ with
  top dead-zone thickness in the Monte-Carlo simulation.  The value in
  the data is shown in blue. The dashed blue lines show the 2\% statistical
  error on the measured value. The red curve is a best fit to an exponential
  function.}
\label{fig:dztune}
\end{center}
\end{figure}

 Figure~\ref{fig:dztune} shows the variation of the ratio $R(\mathrm{multiplet}/59)_{\mathrm{data}}$ 
 with dead layer thickness, after reducing the copper layer. The data measurement with its error
 bar is also shown (see discussion below for our systematic error estimation).
 The best-fit dead layer is $0.760\pm0.025$ mm thick, in good agreement with the manufacturer's
 rough estimate of 700 microns. During this run the side dead layer thickness was kept fixed 
 to its nominal value. The error is statistical only at this stage.

 We placed another Am source (40 kBq, guaranteed with 5\% uncertainty by its vendor) on one side of the 
 crystal, 1 cm from the side of copper endcap. The same methods yields $1.025 \pm 0.025$ mm thickness 
 for the side dead layer, having fixed the top dead layer to the previous best fit.
 
 Iterating the procedure once more shows that the two previously found values are stable and correspond
 to the best fit.
 
 \subsection{Systematic error budget}
\label{sec:systematic}
  Several sources of systematic uncertainty affect our measurements.
We estimate the distance from the source to the endcap with a 1 mm 
uncertainty, and the distance to the center with a 3 mm inaccuracy
(conservatively).
We varied the source position in the simulation by $\pm 3$ mm and used 
the count-rate variation (4.3\%) as the systematic error coming from 
inaccurate distance measurements.
The plastic thickness of the source housing and of the source holder
was estimated with a 0.3 mm accuracy. Using variations of this amount
in the simulation led to a 1.4\% variation of the count-rate.
The high activity of the source causes a dead-time below 10\%. Any systematic
error in the dead-time measurement affects all lines equally, therefore 
this source of error does not contribute to the total error.
Summing up the different contributions in quadrature, we estimate
that our dead zone measurement has a 5\% systematic error. 

\subsection{Tuning of the active volume}

  After tuning the dead layer around the crystal, we took spectra using the following
sources:
\begin{itemize}
 \item $^{22}$Na, $3.66\pm 0.07$ kBq 
 \item $^{60}$Co, $2.81\pm 0.07$ kBq 
 \item $^{137}$Cs, $4.80 \pm 0.09$ kBq 
 \item $^{228}$Th, $20.9 \pm 0.7$ kBq 
\end{itemize}
  
 All sources except the thorium source are from the same manufacturer (LEA-CERCA)
and were purchased and calibrated to with 2\% precision by the manufacturer in March 2009. 
They have the same housing: a plastic cylinder 25 mm high and 9 mm in diameter.
The active area is circular and flat, located 1 mm from one end of the cylinder,
and is 7 mm in diameter. 
 The $^{228}$Th source is a plastic cylinder, 25.4 mm in diameter and 6.35 mm high.
The active area is 5 mm in diameter and 3.18 mm high, encased in the plastic housing.
It was purchased in July 2002 and is guaranteed by the manufacturer to be known to
better than 3.3\%.

We used GEANT4's radionuclide decay module to simulate the relevant gamma ray lines.
For thorium this includes all its daughter nuclei.

All sources were uncollimated. The first three were placed 26.0 cm above the center 
of the detector endcap, and the thorium source was placed 8.0 cm above the detector endcap.
All were placed on the same plastic holder as the AmBe source used at the previous stage.
The sources' plastic housings as well as the holder were simulated.

We compared the full energy peak efficiencies in the data and in the
Monte-Carlo simulations, and found a mean negative bias of about -5\%
in the simulation compared to the data.  This shows that at this
stage, after tuning of the dead zone, our simulation was
underestimating the efficiency of the crystal by a few percent. As
explained by other authors (e.g. \citealt{Hurtado2004764}), this is a
sign that the active volume of the germanium crystal is not well known 
and should also be tuned.  

\begin{figure}[!htb]
\begin{center}
\includegraphics[scale=0.4]{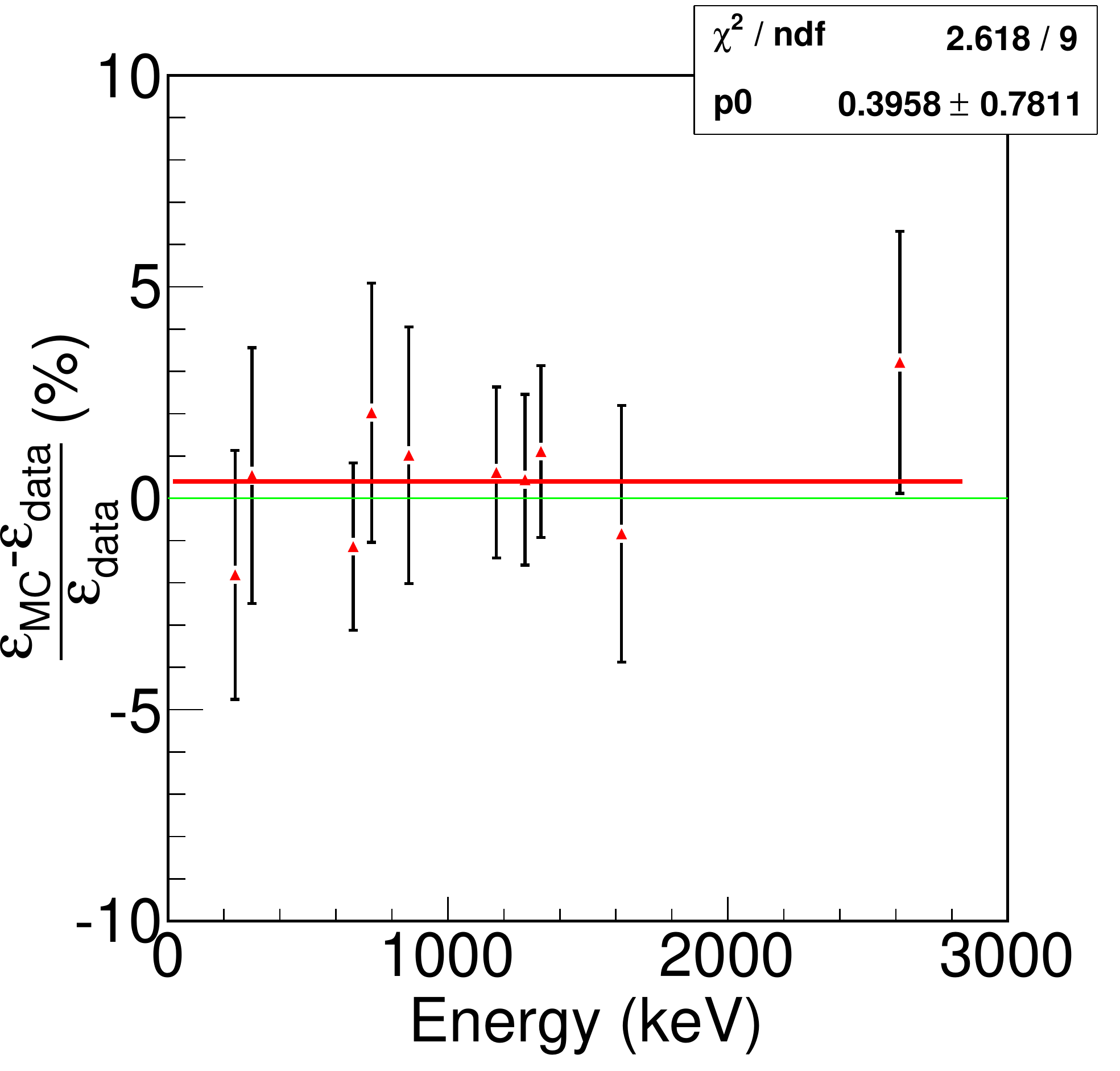}
\caption{Relative data/MC difference after tuning, for several gamma ray lines.
The red line is the best fit to a constant, $0.4\%\pm0.8\%$.}
\label{fig:energy}
\end{center}
\end{figure}

  In order to compensate for the higher observed efficiency, we
increased the active volume in the simulation by reducing the radius of the
borehole and enlarging the radius of the whole detector.  We
found that a tuned borehole radius of 3.0 mm (instead of the nominal
4.5 mm) and a radius of 41.1 mm (instead of the nominal 40.9 mm)
reduced the negative bias to $0.4\%\pm 0.8\%$ (best fit to a
constant), over an energy range covering 300 keV to 2.6 MeV,
sufficient for our radioassay measurements.  The results of the tuning
are shown in figure~\ref{fig:energy}.  With these parameters, the
active volume of the germanium crystal is 362.5 cm$^{3}$.

At this stage, the efficiency of the detector is known to better than 3\% (since all
points are within the $\pm 3\%$ lines), which is sufficient for our needs.

\section{Application to the radioassays of photo-multiplier tubes}

   After tuning the detector's Monte-Carlo simulation to the required accuracy,
we used the large counting chamber inside the shield to radioassay
three photo-multiplier tubes (PMT). One is an 8-inch, Hamamatsu R5912 tube and the
other two are 10-inch, Hamamatsu R7081 tubes.  
They are used respectively in the Nucifer \citep{Nucifer}, Double-Chooz \citep{DChooz}
and Antares \citep{Antares} experiments.  Both Nucifer and Double-Chooz PMTs are
low-background, using specially chosen glass to lower its radioactive
contents, while the Antares PMT is a regular version.  According to
the manufacturer's data sheets both tube sizes have similar
performance.

It is of course possible to perform accurate radioassays of PMT parts,
for example using crushed glass placed in a container with well known
efficiency. However, such large PMTs cost several thousand US dollars
apiece, and are not easy to dismantle. Indeed in many neutrino
experiments PMTs represent a large fraction of the experiment's
budget. The method presented here has the advantage of being
non-destructive and does not affect the PMT's performance.

All PMTs were placed with the curved glass surface laid directly on top of the 
copper endcap, held vertically in place using a small acrylic holder.
A special blank run, with only the acrylic holder inside the shield, was used
in order to perform background subtraction.

A complete geometrical model of the PMT's outer glass surface was implemented using
geometrical code from GLG4sim \citep{glg4sim}, which is used in several neutrino detector simulation codes.
It is well known that especially in large PMTs, glass is the dominant source
of radioactive background.

\begin{figure}[!htb]
\begin{center}
\includegraphics[scale=0.4]{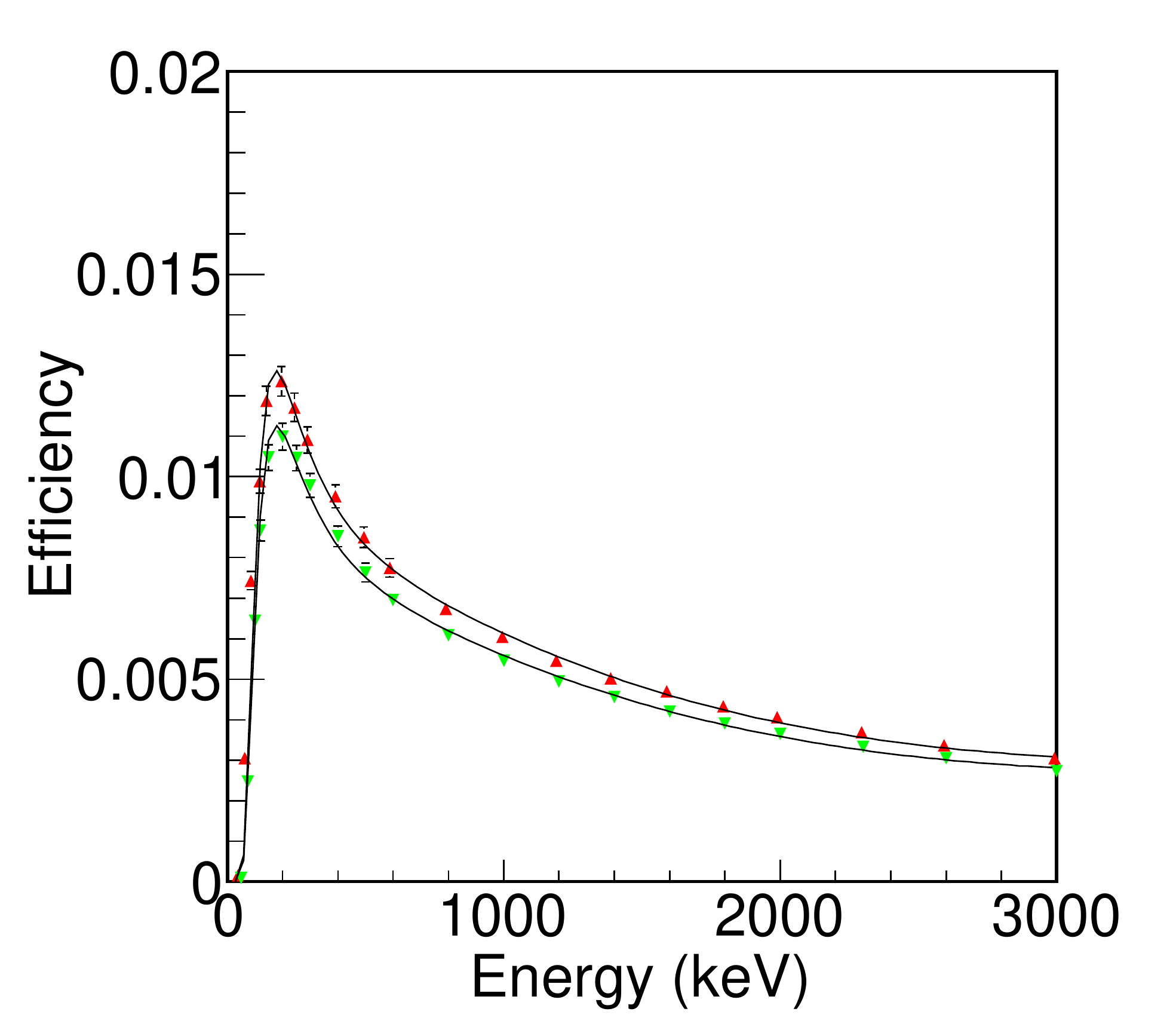}
\caption{Efficiency as a function of energy for 10'' (downward triangles) and 8'' (upward triangles)
PMTs laid vertically on top of the endcap, determined with our tuned simulation, along with best
fits to a $\displaystyle\exp\left(\sum\limits_{i=0}^{5}a_i \log(E)^i\right)$ shape. }
\label{fig:pmteff}
\end{center}
\end{figure}

We assumed that radioelements were uniformly distributed in the PMT glass,
radiating $\gamma$-rays isotropically, and obtained efficiency curves for the experiment, 
which can be seen in figure~\ref{fig:pmteff}.

For each PMT, our peak finding algorithm was able to identify the main lines from natural
radioactivity, namely the U series
($^{226}$Ra, $^{214}$Pb, $^{214}$Bi), the Th series ($^{212}$Pb, $^{208}$Tl),
and natural K. As expected, no other radioelements were found.
 For each line, a 4 keV-wide interval around each peak was used to determine the net count rate,
and eventually the activity. The statistical treatment is that of Currie \citep{currie}.
We assumed that all decay chains were at secular equilibrium.
Tables~\ref{table:pmt-dc},\ref{table:pmt-an} and \ref{table:pmt-nu} summarize 
the measurements for each of the decay series in each PMT.
When no measurement can be made, we report the detection limit at 95\% CL.
The decision threshold is also at 95\% CL.

 The $^{226}$Ra line at 186.2 keV is so close to the 185.9 keV line of $^{235}$U as to
be indistinguishable by our detector. Assuming that PMTs are contaminated with natural
uranium containing 0.072\% of $^{235}$U, it can be determined that 43\% of the counts 
in the 186 keV peak can be attributed to $^{235}$U. 
We have used this correction in order to obtain the $^{238}$U
contents from the 186 keV line in our samples.

Our measurements show that by choosing low background glass,
the K content is reduced by a factor of 7, while the Th content
is lowered by a factor of 2. The observed uranium content in low background glass
depends on the PMT, but is roughly 2.5 times lower than in standard glass.

These measurements can be used in present or future neutrino
experiments as inputs for their background estimations. In experiments
with very stringent background requirements such as Double-Chooz, choosing
low background glass is essential.

\begin{table}[!htbp]
  \begin{tabular}{l|c|c}
    \hline
    \hline
    $\gamma$-lines & Decay chain & Measurement in mBq/kg (fraction)\\
    \hline
    $^{226}$Ra at 186 keV & $^{238}$U & $< 1.7\,10^3$ ($<134$ ppb)\\
    $^{214}$Pb at 295, 352 keV & $^{238}$U & $9.96\,10^{2} \pm 86$ ($81 \pm 7$ ppb)\\
    $^{214}$Bi at 609, 1120, 1765 keV & $^{238}$U & $ 1.32\,10^{3}\pm 90$ ($107 \pm 7$ ppb)\\
    $^{208}$Tl at 2615 keV & $^{232}$Th & $1.42\,10^2 \pm 38$ ($35 \pm 9.4$ ppb)\\
    $^{40}$K at 1460 keV & $^{40}$K & $1.36\,10^3 \pm 40$ ($44.3 \pm 13.1$ ppm)\\
    \hline
    \hline
  \end{tabular}
  \caption{Summary of measurements for the Double-Chooz 10'' PMT, for the most important lines. The results
are in mBq/kg and g/g. The errors are purely statistical, systematics on the efficiency are not
included.}
  \label{table:pmt-dc}
\end{table}

\begin{table}[!htbp]
  \begin{tabular}{l|c|c}
    \hline
    \hline
    $\gamma$-lines & Decay chain & Measurement in mBq/kg (fraction)\\
    \hline
    $^{226}$Ra at 186 keV & $^{238}$U & $3.09\,10^3 \pm 5.32\,10^2$ ($ 250\pm 43$ ppb)\\
    $^{214}$Pb at 295, 352 keV & $^{238}$U & $3.88\,10^{2} \pm 90$ ($314 \pm 7$ ppb)\\
    $^{214}$Bi at 609, 1120, 1765 keV & $^{238}$U & $ 3.72\,10^{3}\pm 92$ ($301\pm 7$ ppb)\\
    $^{208}$Tl at 2615 keV & $^{232}$Th & $3.46\,10^2 \pm 39$ ($85 \pm 10$ ppb)\\
    $^{40}$K at 1460 keV & $^{40}$K & $9.33\,10^3 \pm 4.35\,10^2$ ($304 \pm 14$ ppm)\\
    \hline
    \hline
  \end{tabular}
  \caption{Summary of measurements for the Antares 10'' PMT, for the most important lines. The results
are in mBq/kg and g/g. The errors are purely statistical, systematics on the efficiency are not
included.}
  \label{table:pmt-an}
\end{table}

\begin{table}[!htbp]
  \begin{tabular}{l|c|c}
    \hline
    \hline
    $\gamma$-lines & Decay chain & Measurement in mBq/kg (fraction)\\
    \hline
    $^{226}$Ra at 186 keV & $^{238}$U & $<1.23\,10^3$ ($<100$ ppb)\\
    $^{214}$Pb at 352 keV & $^{238}$U & $5.01\,10^{2} \pm 75$ ($40.5 \pm 6$ ppb)\\
    $^{214}$Bi at 609, 1120, 1765 keV & $^{238}$U & $ 6.68\,10^{2}\pm 67$ ($54.2\pm 5.4$ ppb)\\
    $^{208}$Tl at 2615 keV & $^{232}$Th & $1.68\,10^2 \pm 31$ ($41.5 \pm 7.7$ ppb)\\
    $^{40}$K at 1460 keV & $^{40}$K & $1.27\,10^3 \pm 3.3\,10^2$ ($41.2 \pm 10.7$ ppm)\\
    \hline
    \hline
  \end{tabular}
  \caption{Summary of measurements for the Nucifer 8'' PMT, for the most important lines. The results
are in mBq/kg and g/g. The errors are purely statistical, systematics on the efficiency are not
included.}
  \label{table:pmt-nu}
\end{table}

\section{Conclusion}

 In this work we have presented our low background facility, located near Paris 
in an underground hall formerly used for accelerator physics.
Low background rates of $<0.3$ counts/second/100 cm$^3$ were reached with various
rejection techniques, even with our large counting chamber.
Background rejection could be further improved with
a more efficient active muon veto, covering all sides of the detector, and also
with an improved radon flushing system in the clean room.
We developed a Monte-Carlo simulation for the p-type HPGe detector, and tuned several
geometry parameters to data obtained with point sources, reaching an accuracy on
the efficiency better than 3\% over the relevant energy range.
We then used the simulation to calculate the efficiency of detection for gamma
rays produced inside the glass of large photomultiplier tubes from three neutrino
detection projects, and measured the U, Th and K contamination inside the glass,
with an accuracy of $\sim 10$ ppb on the thorium and uranium contents, and of
$\sim$ 15 ppm on the potassium content.

\section{Acknowledgements}
 We are very grateful to CEA/Saclay IRFU for their loan of the ALS facility 
and for their support of this project, of Double-Chooz and of Nucifer. 
In particular we would like to thank J.-l.~Fallou, A.~Le~Saux and M.~Mur.
We thank the Antares Collaboration for their loan of a 10'' PMT.
We are also grateful to M.~Hofmann (TUM) for sharing his simulation code
in the early stages of this work.
This project was made possible in part thanks to the Regents of the University
of California's financial support.
We would like to thank H.~Simgen (MPIK) for his advice, as well as V.~Durand,
S.~Herv\'{e}, C.~Jeanney and P.~Starzynski for their help with the muon veto.

\end{document}